\documentclass[a4paper]{jpconf}
\usepackage{graphicx}
\usepackage{float}
\newcommand{\GF}{G_{\rm F}}
\newcommand{\CV}{C_{\rm V}}
\newcommand{\CA}{C_{\rm A}}
\newcommand{\bk}{{\bf k}}
\newcommand{\bp}{{\bf p}}
\newcommand{\bq}{{\bf q}}

\renewcommand\({\left(}
\renewcommand\){\right)}
\renewcommand\[{\left[}

\begin{document}
\title{Solar neutrinos at keV energies: thermal flux\footnote{The results quoted here were obtained in reference~\cite{Vitagliano:2017odj} and presented by E. Vitagliano at TAUP 2017.}}

\author{Edoardo Vitagliano,$^1$ Javier Redondo,$^{1,2}$
Georg Raffelt$^1$}
\address{$^1$ Max-Planck-Institut f\"ur Physik (Werner-Heisenberg-Institut), F\"ohringer Ring 6, 80805 M\"unchen, Germany}
\address{$^2$ University of Zaragoza, P. Cerbuna 12, 50009 Zaragoza, Spain}
\ead{edovita@mpp.mpg.de}

\begin{abstract}
The neutrino flux at Earth is dominated in the keV energy range by the neutrinos produced in the Sun through thermal processes, namely photo production, bremsstrahlung, plasmon decay, and emission in free-bound and bound-bound transitions of partially ionized elements heavier than hydrogen and helium. Such a component of the neutrino flux is conspicuously absent from popular analyses of the all-sources spectrum at Earth, whereas if detected it could be a source of information about solar physics. Moreover, it would be the relevant background for keV-mass sterile neutrino dark matter direct searches. \end{abstract}

\section{Introduction: what is the thermal neutrino flux, and why is it important?}
The Sun is commonly recognized as a neutrino factory, being the most important source of neutrinos in the MeV range because of the nuclear reactions happening in its core \cite{Spiering:2012xe}. However, it also produces neutrinos through the so-called thermal processes. These reactions are triggered by the electromagnetic interaction of the particles constituting the plasma (electron, ions, photons), and produce a flux which is the dominant component in the keV energy range.
Unlike other stars, the energy loss due to thermal neutrinos in the Sun is small; nevertheless, the flux caused by the latter is important in different regards. On the one hand, it is a window into solar physics, bringing information about the metallicity of the Sun. On the other hand, the thermal flux is also important because it would be the background in direct detection experiments searching for the keV-mass sterile neutrino; the latter is a well motivated dark matter candidate. Its mass is big enough to evade the Tremaine-Gunn bound yet small enough to potentially help solving small scale controversies such as the cusp-core problem~\cite{Adhikari:2016bei}. Moreover, its existence would obviously be of paramount importance in elementary particle physics, because the right handed neutrino is a missing ingredient in the Standard Model updated to the fact that neutrinos have mass. Some recently proposed experiments include ideas such as using dysprosium targets \cite{Lasserre:2016eot}.

Our analysis improves the only detailed previous calculation of the thermal flux~\cite{Haxton:2000xb} in different regards. Most importantly, we include the very relevant bremmstrahlung process, correct a spurious resonance and include free-bound and bound-bound processes by taking advantage of detailed photon opacity calculations, inspired by the recent axion literature~\cite{Redondo:2013lna,Redondo:2013wwa}. The detailed analysis is presented in reference~\cite{Vitagliano:2017odj}; this manuscript, in which our findings are summarized, is organized as follows: in section~\ref{secabcd} we review the processes contributing to the thermal flux; in section~\ref{secflux} we show the total flux produced and we include flavor mixing; finally in section~\ref{sec:discussion} the conclusions are reported.

\section{How to compute the thermal neutrino flux: ABCD processes}\label{secabcd}
\begin{figure}\label{fig:processes}
\centering
\includegraphics[width=35pc]{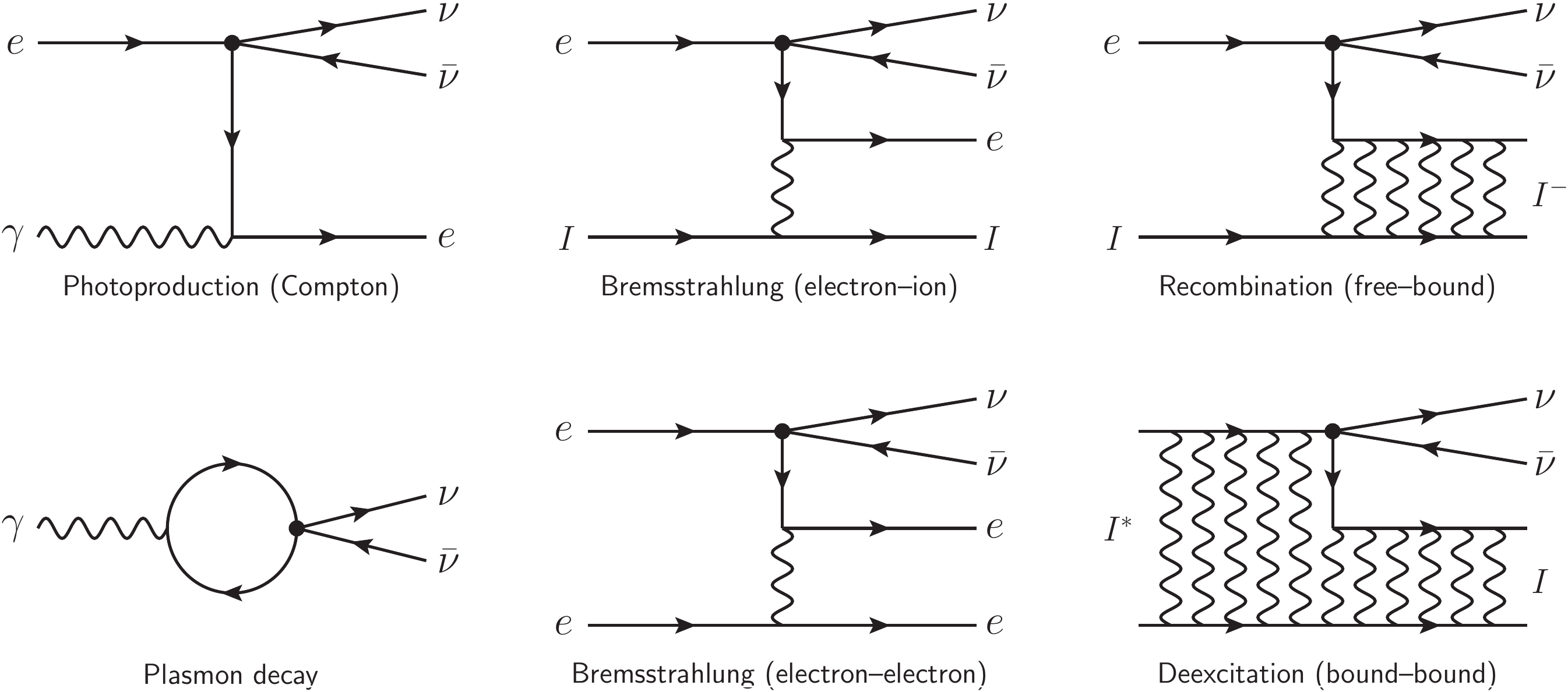}
\caption{Processes contributing to the thermal neutrino flux; figure from \cite{Vitagliano:2017odj} with permission of the authors.}
\end{figure}
Low-energy neutrinos are produced in the solar plasma by pair-production
processes, where nonrelativistic
electrons, whose velocities and spins are coupled to the
ambient electromagnetic fields, are the sources for neutrino pair emission. At
low energies, the relevant weak interactions are well described by an
effective four-fermion local interaction
\begin{equation}\label{eq:NC-interaction}
  {\mathcal L}_{\rm int}=\frac{\GF}{\sqrt{2}}\,
  \bar\psi_e\gamma^\mu(\CV-\CA\gamma_5 )\psi_e\,
  \bar\psi_\nu\gamma_\mu(1-\gamma_5)\psi_\nu\,.
\end{equation} The effective coupling constants
for the vector and axial-vector interaction, $\CV$ and $\CA$, are different
for $\nu_e$ and the other flavors because of the possibility of the former to be produced through a $W$ boson, so that \begin{equation}
   \CV^2=0.9263~~\hbox{for}~\nu_e\bar\nu_e
   \quad\hbox{and}\quad
   \CV^2=0.0014~~\hbox{for}~\nu_{\mu,\tau}\bar\nu_{\mu,\tau}\,;
\end{equation}
thus the emitted flux is not equally distributed among the different flavors. The vector (res. axial) current is the source for electric (magnetic) dipole radiation caused by the time variation of the electron
velocity (spin). However in the nonrelativistic
limit there is no term due to interference, so any process implies a rate
proportional to $(a\, \CV^2+b\,\CA^2)\GF^2$ with different coefficients $a$ and $b$ for different transitions; we will show an example below. The structure of the emission rates allows us to compare them to those for axions (axial current interaction) or hidden photons (vector current interaction), as well as to photon absorption rates.
The various neutrino flux contributions, shown in figure~1, can also be listed in a mnemonically helpful way as ABCD processes:
\begin{itemize} 
\item \textbf{A}tomic deexcitation, including free-bound (fb, also known as electron capture or recombination) and bound-bound (bb) processes;
\item \textbf{B}remsstrahlung, including free-free (ff, atomic bremsstrahlung) and electron-electron (ee) interactions;
\item \textbf{C}ompton scattering;
\item Plasmon \textbf{d}ecay, i.e., the decay of a photon, which is kinematically allowed thanks to its dispersion relation in a medium.
\end{itemize}
As a paradigmatic example of the way one can compute the various contributions to the thermal flux, let us consider the bremsstrahlung process. The emission rate of neutrino pairs per unit of volume and time is
\begin{equation}\label{eq:brems-emission-1}
\dot n_{\nu}=n_Z \int\frac{d^3\bp_1}{(2\pi)^3}\frac{d^3\bp_2}{(2\pi)^3}
\frac{d^3\bk_1}{(2\pi)^3}\frac{d^3\bk_2}{(2\pi)^3}\,f_1(1-f_2)\,
\frac{\sum_{s_1,s_2}|\mathcal{M}|^2}{(2m_e)^2 2\omega_1 2\omega_2}\,
2\pi\delta(E_1-E_2-\omega)\,,
\end{equation}
where the sum is taken over the electron spins and $f_1(\bp_1)$ and $f_2(\bp_2)$ are the initial and
final-state electron occupation numbers, while $(\omega_1,\bk_1)$ and $(\omega_2,\bk_2)$ are the neutrino four-momenta. Here microscopic physics enters through the squared amplitude and thermal physics enters through the electron distribution function; the latter introduce an unavoidable uncertainty of about 10\% due to the choice of a specific solar model. Note that the delta function does not depend on the momentum transfer, because the mass of the ions turns out to be much bigger than other energy scales; this is known as ``long wavelength approximation''.

The latter allows us to express the neutrino pair emission rate of equation~(\ref{eq:brems-emission-1}) factorizing the term related to emitted radiation (the neutrino pairs) from the term describing the medium (thermal electrons
interacting with nuclei)
\begin{equation}\label{eq:brems-emission-2}
\dot n_{\nu}=n_Z n_e \frac{8\,Z^2 \alpha^2}{3}
\int\frac{d^3\bk_1}{2\omega_1(2\pi)^3}\frac{d^3\bk_2}{2\omega_2(2\pi)^3}
\(\frac{\GF}{\sqrt{2}}\)^2
\frac{\(\CV^2 \bar M_{\rm V}^{\mu\nu}+\CA^2 \bar M_{\rm A}^{\mu\nu}\)N_{\mu\nu}}{\omega^4}
\,{\cal S}(\omega)\, ,
\end{equation}
with $\omega=\omega_1+\omega_2$ being the total energy of the neutrino pair. Moreover, we have used the possibility of factorizing the squared amplitude in a medium dependent tensor $\bar M$ and a radiation (in our case, neutrinos) tensor $N$, $\sum_{s_1,s_2}|\mathcal{M}|^2\propto \(\CV^2 \bar M_{\rm V}^{\mu\nu}+\CA^2 \bar M_{\rm A}^{\mu\nu}\)N_{\mu\nu}$. The term describing the medium statistical properties, known as response function, is
\begin{equation}\label{eq:brems-response-1}
{\cal S}(\omega)=\frac{(4\pi)^2}{(2m_e)^2}\frac{1}{n_e}
\int\frac{d^3\bp_1}{(2\pi)^3}\frac{d^3\bp_2}{(2\pi)^3}\,f_1(1-f_2)\,\frac{1}{\bq^2}\,
2\pi\delta(E_1-E_2-\omega)\,,
\end{equation}
where $\bq=\bp_1-\bp_2$ is the momentum transfer in the electron-nucleus collision
mediated by the Coulomb field. Collective effects due to the presence of the plasma can modify $\cal S$; we have discussed such corrections, affecting several different processes and accounting for a 20\% modification of the flux, in reference~\cite{Vitagliano:2017odj}.

\begin{figure}[b!]
\centering
\includegraphics[height=3.8cm]{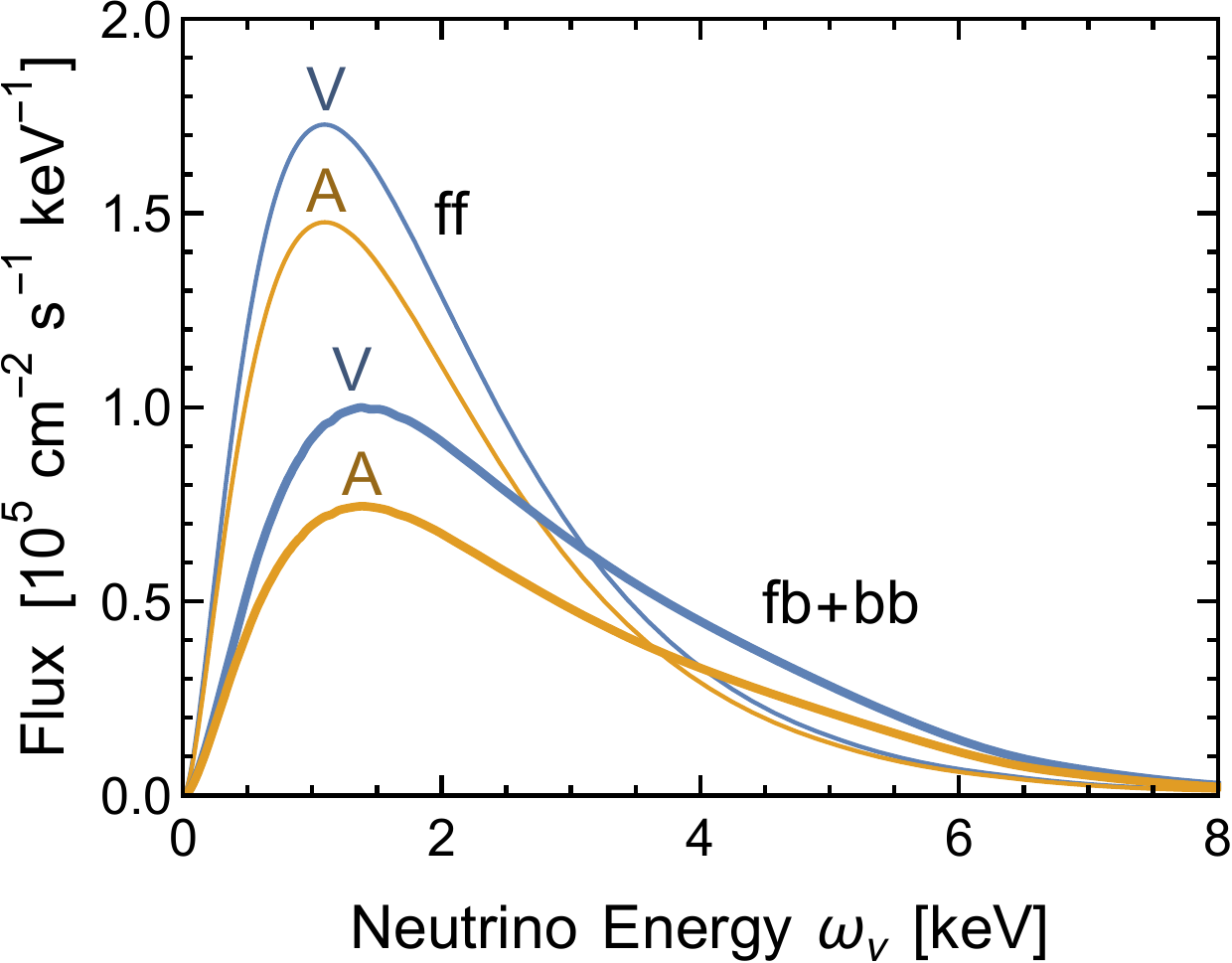}
\includegraphics[height=4.5cm]{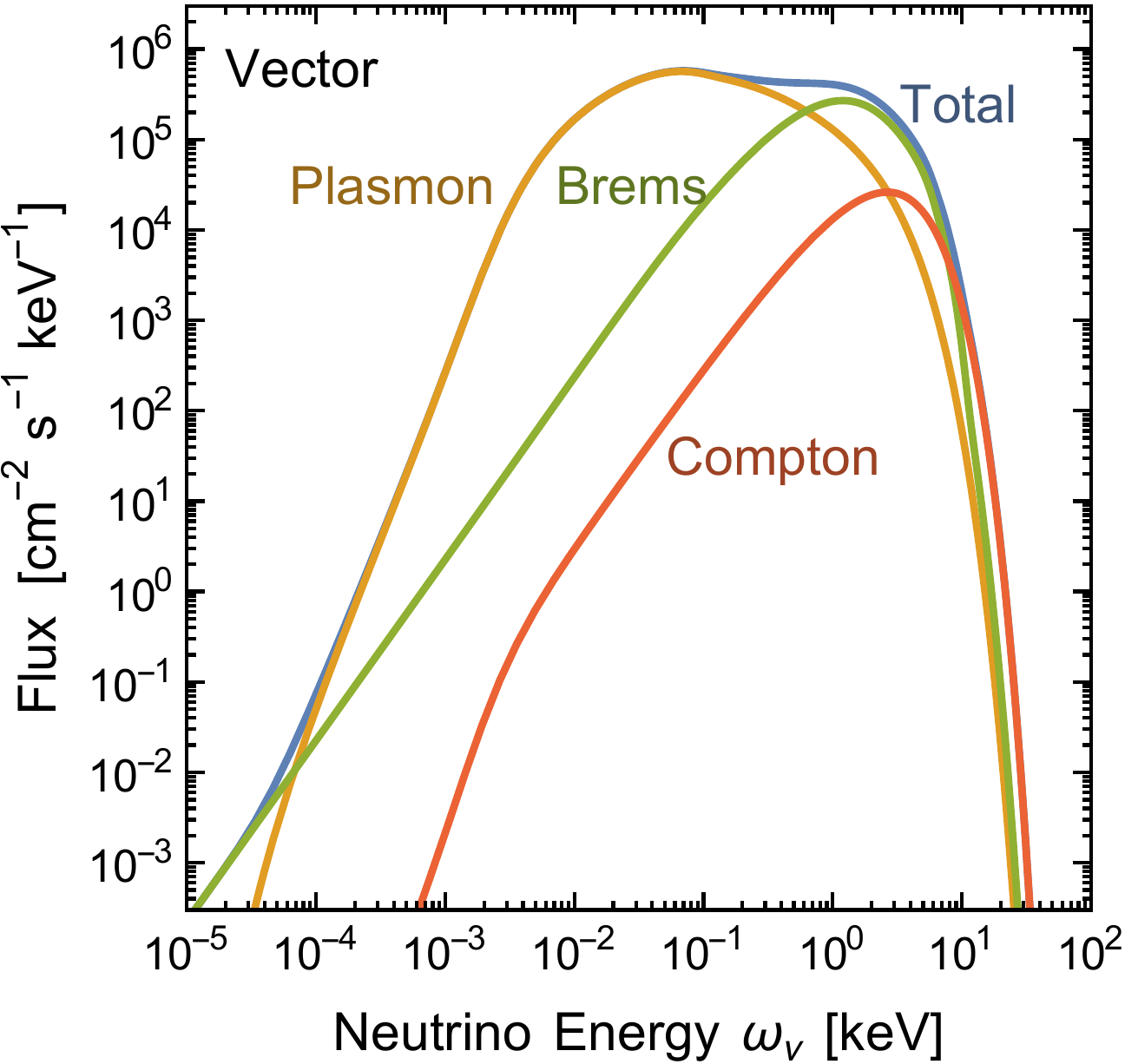}
\caption{{\em Left panel:} Solar neutrino flux at Earth from free-free (ff), free-bound (fb) and
  bound-bound (bb) electron-ion transitions for the vector (V) and axial-vector (A)
  contributions. For proper fluxes, multiply
  with $\CV^2$ and $2\CA^2$, respectively. {\em Right panel:} Spectrum at Sun produced by vector coupling, for proper
  flux multiply with $\CV^2$. Plot from \cite{Vitagliano:2017odj} with permission of the authors.}
\label{fig:awef}
\end{figure}

Integration over neutrino emission angles gives the
neutrino emission spectrum by using $\omega_1=\omega_\nu$ and
$\omega_2=\omega-\omega_\nu$. We integrate over the anti-neutrino
energy to obtain the final expression
\begin{eqnarray}\label{eq:brems-emission-3}
  \frac{d\dot n_{\nu}}{d\omega_\nu}&=&n_Z n_e\,\frac{8\,Z^2 \alpha^2}{3}\,
  \(\frac{\GF}{\sqrt{2}}\)^2\frac{1}{3\pi^4}
  \int_{\omega_\nu}^\infty d\omega\,{\cal S}(\omega)\,
  \frac{\omega_\nu^2(\omega-\omega_\nu)^2}{\omega^4}\nonumber\\[2ex]
&&\kern5em{}\times
  \Bigl[\CV^2\(3\omega^2-2\omega\omega_\nu+2\omega_\nu^2\)
  +2\CA^2\(3\omega^2-5\omega\omega_\nu+5\omega_\nu^2\)\Bigr]\,,
\end{eqnarray}
which is the rate of (anti)neutrino emission per unit volume, unit time,
and unit energy interval. 

We have calculated the free-bound and bound-bound contribution exploiting the relation between neutrino pair emissivity and photon opacity. This relation is straightforwardly obtained in the Bremsstrahlung case, where
\begin{equation}\label{eq:brems-photon-emission-2}
  \frac{d\dot n_{\gamma}}{d\omega}=n_Z n_e\, \frac{8\,Z^2 \alpha^2}{3}\,
  \frac{\alpha}{\pi}\,\frac{{\cal S}(\omega)}{\omega}\,.
\end{equation}
By inverting equation~(\ref{eq:brems-photon-emission-2}) and exploiting the relation between emission and absorption rates due to thermal equilibrium
\begin{equation}
  \frac{d\dot n_{\gamma}}{d\omega}\big|_{\rm abs} =e^{\omega/T}\frac{d\dot n_{\gamma}}{d\omega}\, ,
\end{equation}
we obtain the response function in terms of the photon opacity. Photon opacities used in stellar structure calculations include already absorption in bound-free and bound-bound transitions by including electron level occupation probabilities and transition matrix elements for the relevant ions; an equivalent calculation from the scratch would require an enormous effort. The latter processes bring information about solar metallicity and are dominating at around 5 keV (figure~\ref{fig:awef}).

\section{Solar neutrino flux at Earth}\label{secflux}

To obtain the flux as it would be measured at Earth, we have to account for flavor oscillations. As is well known, neutrinos are produced by weak interactions as mixtures of three different mass eigenstates. Considering the long distance between the Sun and our planet, one can safely consider the neutrinos arriving at Earth as a flux of mass eigenstates.

The axial-current processes flux is straightforwardly obtained. Indeed, $\CA^2$ is the same for electron and other flavor neutrinos, so we can read the flux of different mass-eigenstates by simply taking the spectrum as it is produced in the Sun.

On the other hand, vector-current processes produce almost exclusively electron neutrinos. One then needs to account for flavor oscillations. For neutrino energies in the keV range, the matter effect is negligible, so that we can treat the oscillations as if they were happening in vacuum. The scale of distances, with the Sun being much bigger than the oscillation length, and the Sun-Earth distance being bigger than both, allows us to consider the spectrum an incoherent mixture of mass eigenstates. 
The final fluxes in terms of mass eigenstates are shown in figure~\ref{fig:summaryflux}.

\section{Discussion and summary}\label{sec:discussion}
We have computed the neutrino flux produced in the Sun through thermal processes, the dominant component of the flux at Earth in the keV energy range. Compared to the previous study~\cite{Haxton:2000xb}, we have included the bremsstrahlung processes, and we got rid of a spurious contribution (the
\begin{figure}
\centering
{\includegraphics[height=4.5cm]{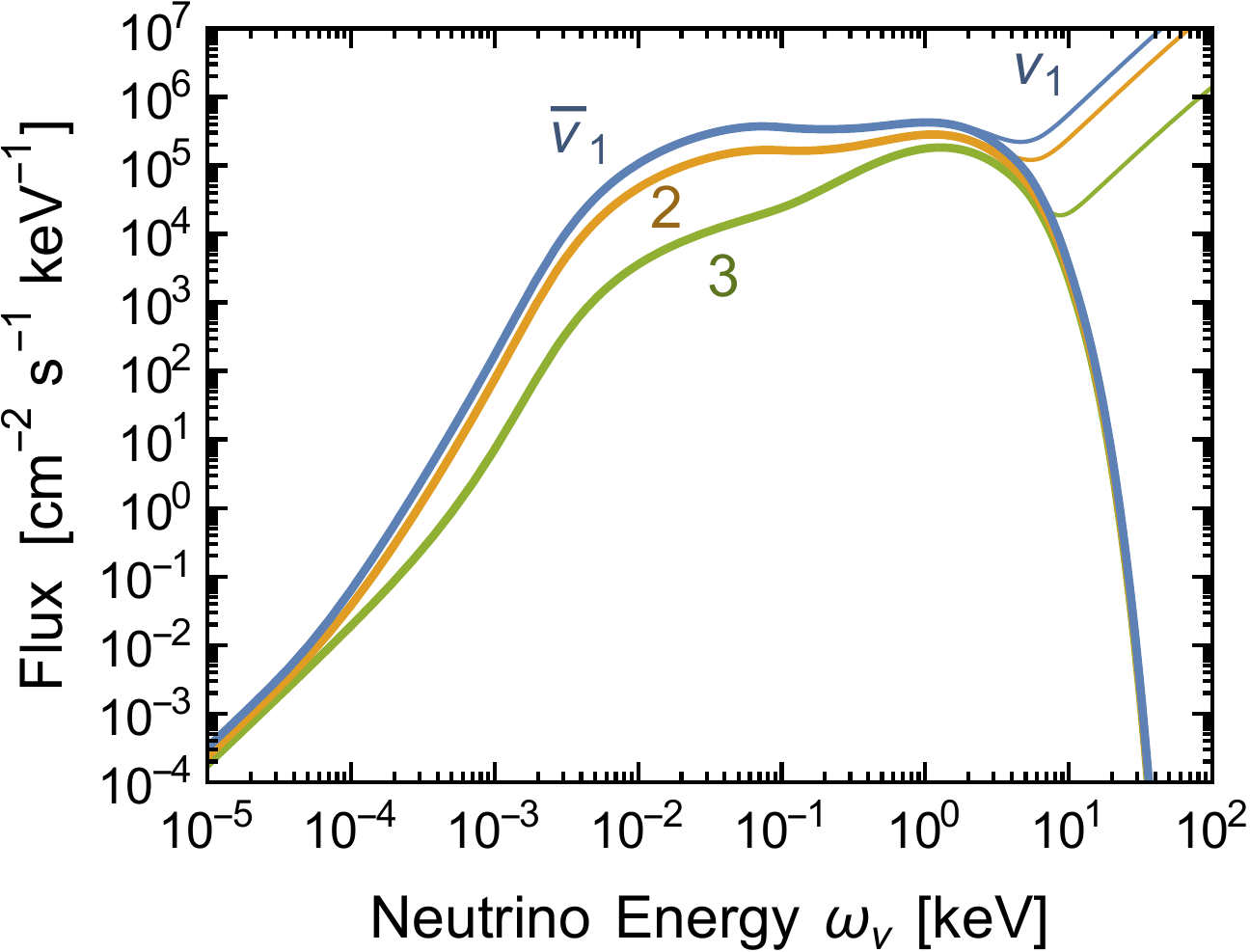}
\includegraphics[height=4.5cm]{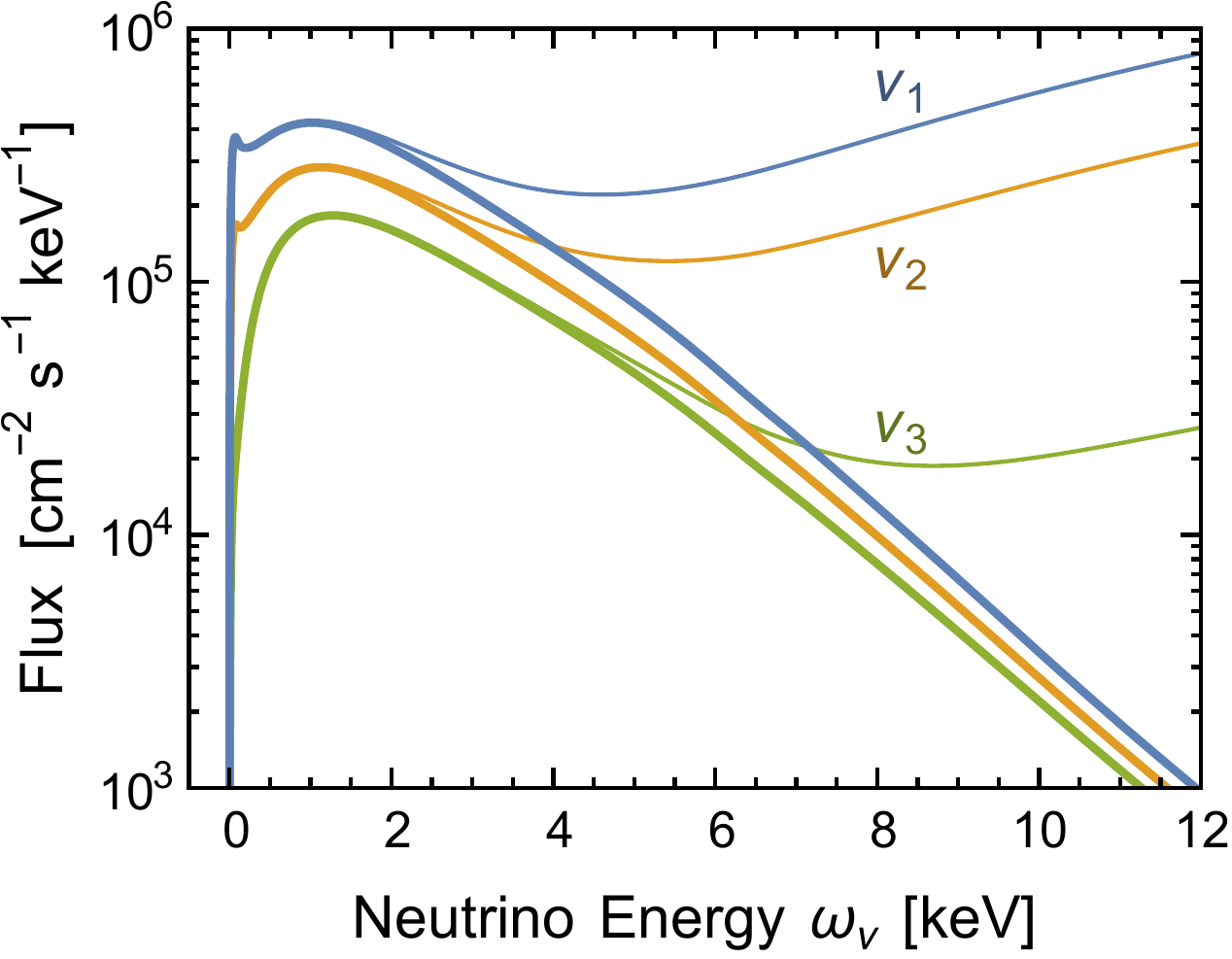}}
\vskip12pt
{\includegraphics[height=4.5cm]{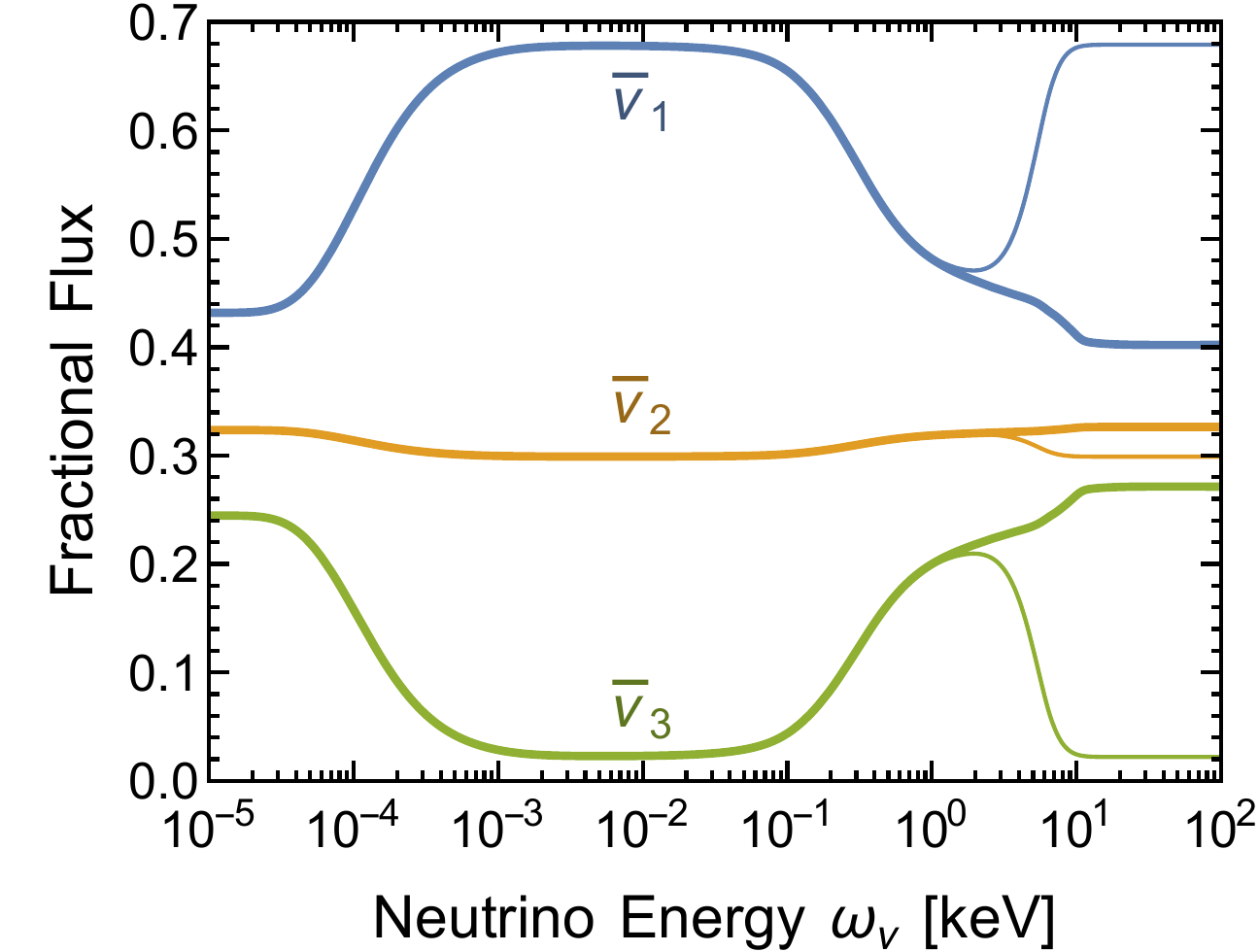}
\includegraphics[height=4.5cm]{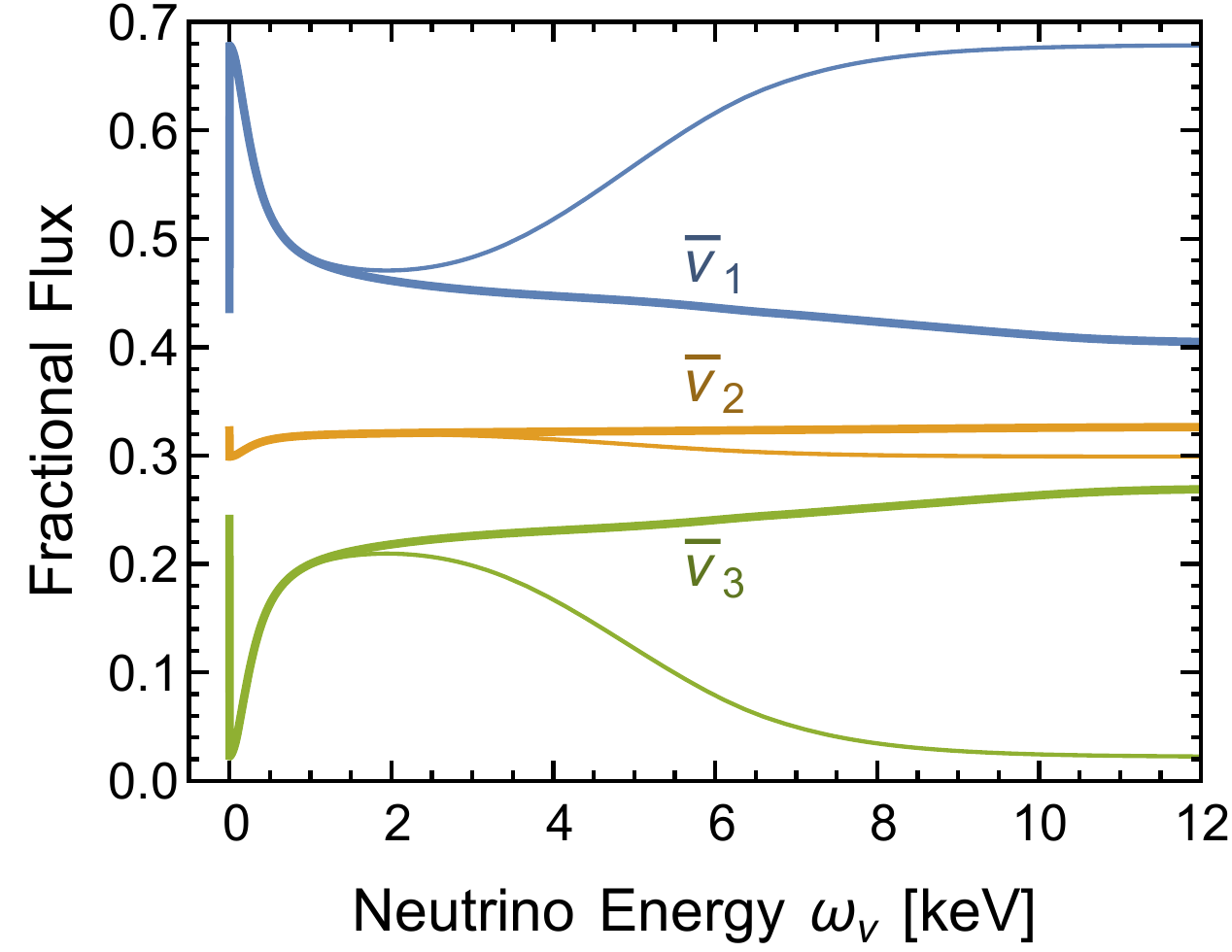}}
\caption{Solar neutrino fluxes of different mass eigenstates at Earth in the keV range.  Thick lines are for $\bar\nu$,
  thin lines for $\nu$ which includes a contribution of $\nu_e$ from the nuclear
  pp reaction. The other
  source channels are thermal processes which produce both $\nu$ and
  $\bar\nu$.  The bottom panels show the fractions of
  the total flux provided by the individual mass eigenstates. Plots from \cite{Vitagliano:2017odj} with permission of the authors.}
\label{fig:summaryflux}
\end{figure}
``plasmon pole'' contribution) which overshadowed other processes; moreover, we improved the free-bound and bound-bound calculation by using an opacity code.

The final result is shown in figure~\ref{fig:summaryflux}. The reasons for an experimental effort to detect keV neutrinos would be manifold: the signal of today is the background of tomorrow. First, this flux brings information about metallicity of the Sun. This feature has already been stressed in reference~\cite{Haxton:2000xb}, but it was stated that such an observation would have been impossible because of the plasmon pole contribution. Moreover, such a detector would be capable of directly detecting a promising dark matter candidate, the keV-mass sterile neutrino. The flux we have computed would be the background to be overcome to detect such a sterile neutrino. We think that our analysis will be useful to different communities, and hope it will help to shed light on both solar physics and fundamental particle physics.

\ack{Thanks go to Alexander Millar for reading the manuscript. 
E.V. is grateful to the TAUP 2017 organizers for kind hospitality and acknowledges funding from the European Union through Grant No. H2020-MSCA-ITN-2015/674896 (Innovative Training Network ``Elusives'').
}


\section*{References}
\begin{thebibliography}{99}
\bibitem{Vitagliano:2017odj}
  Vitagliano E, Redondo J and Raffelt G 2017
  \it{Preprint} \normalfont arXiv:1708.02248
  
\bibitem{Spiering:2012xe}
  Spiering C 2012
  \normalfont
 \it Eur.\ Phys.\ J.  H \normalfont {\bf 37} 515

\bibitem{Itoh:1996}
  Itoh N, Hayashi H, Nishikawa A, Kohyama Y 1996
  \normalfont
 \it Astrophys.\ J.\ Suppl.\ \normalfont {\bf 102}  411

\bibitem{Adhikari:2016bei}
Drewes M \it{et al} \normalfont 2017
  \normalfont
\it  JCAP \normalfont {\bf 1701} 025

\bibitem{Lasserre:2016eot}
Lasserre T, Altenmueller K, Cribier M, Merle A, Mertens S and Vivier M 2016
  \normalfont
  \it{Preprint} \normalfont arXiv:1609.04671



\bibitem{Haxton:2000xb}
  Haxton W and Lin W 2000
  \normalfont
\it  Phys.\ Lett.\ B  \normalfont  {\bf 486} 263

\bibitem{Redondo:2013lna}
  Redondo J and Raffelt G 2013
  \normalfont
\it  JCAP \normalfont  {\bf 1308} 034

\bibitem{Redondo:2013wwa}
  Redondo J 2013
  \normalfont
\it  JCAP \normalfont {\bf 1312} 008



\end{thebibliography}
\end{document}